\documentclass[pmlr,twocolumn]{jmlr} 

\usepackage{booktabs}
\usepackage[load-configurations=version-1]{siunitx} 

\usepackage{caption}
\allowdisplaybreaks

\theorembodyfont{\upshape}
\theoremheaderfont{\scshape}
\theorempostheader{:}
\theoremsep{\newline}

\jmlrvolume{}
\jmlryear{2020}
\jmlrsubmitted{}
\jmlrpublished{}
\jmlrworkshop{Machine Learning for Health (ML4H) 2020} 

\title[EEG-GCNN]{EEG-GCNN: Augmenting Electroencephalogram-based Neurological Disease Diagnosis using a Domain-guided Graph Convolutional Neural Network}

\author{%
\Name{Neeraj Wagh} \Email{nwagh2@illinois.edu}\\
\addr Department of Bioengineering, University of Illinois at Urbana-Champaign
\AND
\Name{Yogatheesan Varatharajah} \Email{varatha2@illinois.edu}\\
\addr Department of Bioengineering, University of Illinois at Urbana-Champaign
}

\begin{document}

\maketitle

\begin{abstract}

This paper presents a novel graph convolutional neural network (GCNN)-based approach for improving the diagnosis of neurological diseases using scalp-electroencephalograms (EEGs). Although EEG is one of the main tests used for neurological-disease diagnosis, the sensitivity of EEG-based expert visual diagnosis remains at $\sim$50\%. This indicates a clear need for advanced methodology to reduce the false negative rate in detecting abnormal scalp-EEGs. In that context, we focus on the problem of distinguishing the abnormal scalp EEGs of patients with neurological diseases, which were originally classified as 'normal' by experts, from the scalp EEGs of healthy individuals. The contributions of this paper are three-fold: 1) we present EEG-GCNN, a novel GCNN model for EEG data that captures both the spatial and functional connectivity between the scalp electrodes, 2) using EEG-GCNN, we perform the first large-scale evaluation of the aforementioned hypothesis, and 3) using two large scalp-EEG databases, we demonstrate that EEG-GCNN significantly outperforms the human baseline and classical machine learning (ML) baselines, with an AUC of 0.90.
\end{abstract}


\section{Introduction}


Neurological disorders (NDs) are diseases of the nervous system involving the brain, spinal cord, nerves, and muscles, and affect $\sim$1 billion people worldwide \citep{world2006neurological}. EEG is one of the main diagnostic tests in neurology, where the visual identification of abnormal brain activity in a brief scalp EEG recording session (20--60 minutes) indicates the potential for NDs. However, it is very common to record EEGs that do not contain visible abnormalities; for example, 50\% of the EEGs recorded from patients with seizures are deemed ``normal'' based on expert visual review \citep{Smithii2}. Such scenarios can cause delays in delivering clinical care and put patients at continued risk for injuries and comorbidities \citep{bouma2016diagnostic}. Thus, there is a critical need to develop EEG-based analytical tools that can enable a more rapid diagnosis of NDs. 


The task of visually identifying abnormal EEG is challenging due to multiple reasons: abnormal discharges may not occur during a short EEG session; they may originate in deeper brain structures like the cingulate, hippocampus, or insula; they may be activated only during sleep, which was not recorded; they may involve too small an amount of cortex to be measurable on the scalp; or subtle abnormalities are highly likely to evade inspection by the naked eye \citep{ebersole1983evaluation}. However, a recent study showed that deviations in EEG spectral characteristics, such as the alpha rhythm, can help distinguish epilepsy patients from healthy individuals even when visual classification was not possible \citep{varatharajah2020electrophysiological}. The study also showed that the spatial patterns of the abnormalities are localized to specific brain regions impacted by the disease. Motivated by those findings, we sought to perform a large-scale evaluation of distinguishing ``normal" EEGs of patients with NDs from EEGs of healthy individuals using state-of-the-art machine learning (ML) techniques.

In this paper, we present EEG-GCNN, a graph convolutional neural network (GCNN)-based approach that achieves state-of-the-art performance in classifying ``normal" EEGs of patients with NDs versus EEGs of healthy individuals. The key contributions of our paper are the following: 1) we present the first large scale evaluation of the task at hand using the EEGs of 208 healthy individuals and 1,385 patients with NDs; 2) the proposed model includes a novel graph representation for EEG data using spatial and functional connectivity measures; and 3) EEG-GCNN achieves an AUC of 0.90 on the held-out test set and significantly outperforms human and classical ML baselines (10\% improvement).

\section{Related Work}


\textbf{Clinical relevance:} Prior research related to the diagnosis of epilepsy has focused on visual identification of common epileptic abnormalities, such as interictal spikes and sharp waves \citep{hauser1982seizure}. Furthermore, the majority of the existing ML-based approaches have targeted automating the visual identification of normal and abnormal EEGs, using expert labels as ground truth \citep{schirrmeister2017deep,roy2018deep,alhussein2019eeg}. However, visual identification of abnormal EEGs is $\sim50\%$ sensitive, and therefore does not provide reliable ground truth labels. A clinically more important question is whether ML can distinguish between healthy EEGs and EEGs of patients that do not contain any visually identifiable abnormalities. A study by \citet{varatharajah2020electrophysiological} provides strong evidence for this hypothesis. They found that deviations in brain health markers like the alpha rhythm help distinguish patients from healthy individuals using a modest sample and hand-tuned features. However, a large-scale evaluation of this hypothesis using state-of-the-art ML approaches has not been performed.

\textbf{Technical relevance:} Prior studies have proposed graphical-model-based approaches to encode the rich spatial and temporal information content of EEG data \citep{varatharajah2017eeg}. However, those studies have hard-coded the spatio-temporal relationships based on domain knowledge and therefore, lack the ability to learn from data. Recent studies have addressed this limitation using GCNNs, albeit being limited to the application of emotion recognition \citep{song2018eeg,wang2018eeg}. They have proposed a dynamical graph convolutional neural network model (DGCNN) with an adjacency matrix reflecting the functional coupling between EEG electrodes. While EEG-GCNN and DGCNN share some commonalities, there are notable differences. First, DGCNN does not fully exploit the power of graph representation as it tries to binarize the adjacency matrix during training. Whereas, EEG-GCNN captures brain connectivity through edge weights on a fully-connected graph. Second, DGCNN makes extensive use of hand-tuned features, whereas, EEG-GCNN applies minimal feature engineering. Finally, based on evidence that the pathological changes induced by chronic neurological diseases are spatially related and show temporally localized functional patterns \citep{hyun2011effects}, we employ a connectivity measure reflecting both the spatial and functional relationships between brain regions.

\section{Data}

We support our experiments by pooling together two publicly available large scalp EEG databases: 1) the Temple University Hospital EEG (TUH EEG) Corpus (\cite{obeid2016temple}), which contains clinical EEG recordings of patients with NDs and 2) the Max Planck Institute Leipzig Mind-Brain-Body (MPI LEMON) Dataset (\cite{babayan2019mind}), which contains resting-state recordings from healthy participants.

\noindent \textbf{TUH EEG:} This dataset comprises of $>$30,000 EEG recordings collected at TUH starting from 2002. The recordings vary in terms of patient ages, diagnoses, medications, channel configurations, and sampling frequencies. A subset of recordings in TUH EEG have been broadly annotated by experts as either ``normal" or ``abnormal", and have been released as a derived dataset called the TUH EEG Abnormal Corpus (TUAB). For our experiments, we only utilize the TUAB recordings that are annotated as ``normal" while ignoring those labeled as ``abnormal", leading to a total of 1385 EEGs from 1385 distinct patients.

\noindent \textbf{MPI LEMON:} This dataset represents a cross-sectional sample of 228 healthy individuals from Leipzig, Germany. The sample comprised two age groups: young adults (ages 20-35) and older adults (ages 59-77). EEG recordings were made using 62 electrodes in the 10-10 sensor configuration with a sampling rate of 2500Hz, for a total of 216 participants. Each subject's session is made up of 16 trials, each 60 seconds long: 8 eyes-closed and 8 eyes-open. We included data from both trials in our experiments. The raw data were corrupted for 8 subjects, which resulted in a useful set of healthy EEGs from a total of 208 healthy subjects.




\section{Data Preprocessing \& Feature Engineering}

%

\begin{figure*}[htbp]
\floatconts
  {fig:feature_extraction}
  {\caption{Schematic representation of the feature extraction process.}}
  {\centering\includegraphics[width=\linewidth]{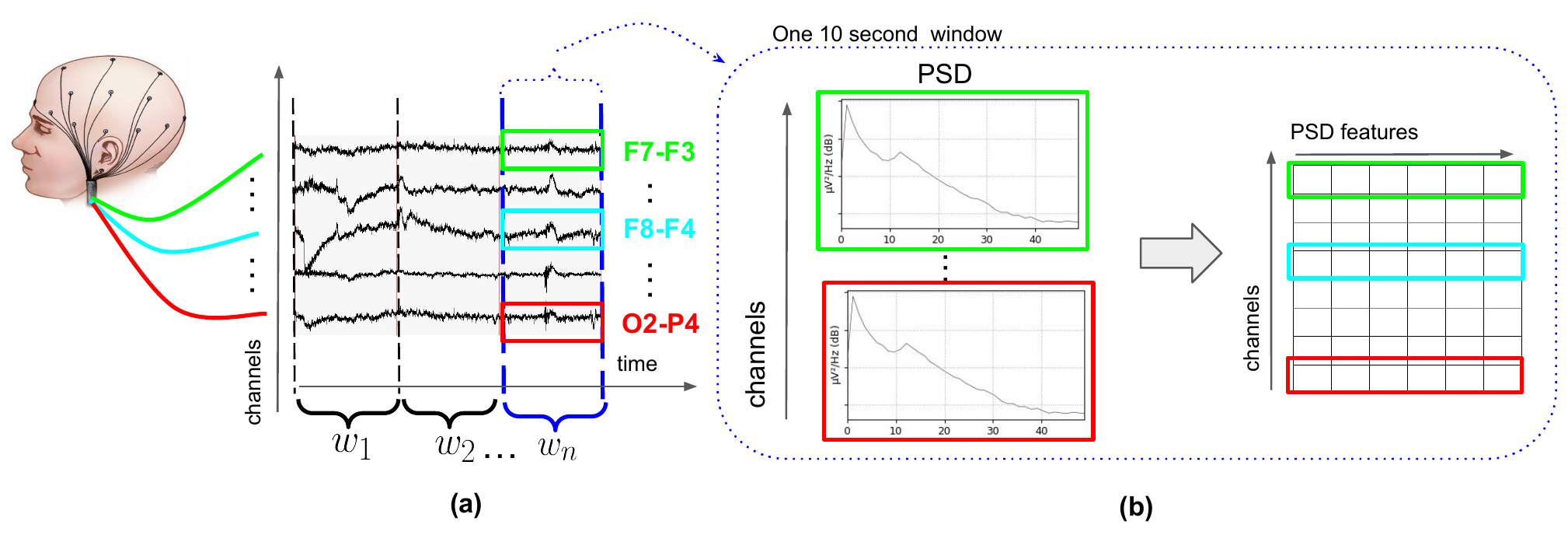}}
\end{figure*}



The overall flow of EEG preprocessing and feature extraction is depicted in Figure \ref{fig:feature_extraction}. 

\noindent \textbf{Preprocessing:} We employed a very minimal level of preprocessing, with each raw recording being transformed as follows: (1) a common subset of bipolar montage electrodes were selected from the raw channels, (2) the recording was resampled to 250Hz, followed by (3) a highpass filter at 1Hz, and finally (4) a notch filter at the power-line frequency of 50Hz. We emphasize that neither were routine physiological EEG artifacts like eye blinks or muscle movements explicitly suppressed nor were bad channels rejected. Implementation was done using routines provided by the MNE-python library (\cite{gramfort2013meg}).

\noindent \textbf{Channel selection:} Because the EEG data of healthy individuals were recorded using a 62-channel 10-10 system, we selected a subset of channels that matched the 10-20 system used to record the EEG data from epilepsy patients. Within the EEG data of selected channels from the healthy and patient populations, we selected 4 bipolar pairs of electrodes from each hemisphere, producing 8 channels of EEG data for each participant. Thus, our analysis involves the following bipolar channels: F7-F3, F8-F4, T7-C3, T8-C4, P7-P3, P8-P4, O1-P3, and O2-P4.   

\noindent \textbf{Windowing:} We divided the preprocessed recordings into contiguous non-overlapping windows of 10s each. Each window consists of an EEG recording from eight bipolar channels as defined earlier. We make a simplifying assumption that the signal in each 10s window is independent of other windows in the same recording. It is important to note that while ML training is done using windows, window predictions are aggregated to form subject predictions. This setup is illustrated in Figure \ref{fig:feature_extraction}(a).

\noindent \textbf{Classification data:} We labeled TUAB recordings as ``diseased" (patient), and MPI LEMON as ``healthy". Each window for a subject was assigned the same label as the parent recording. This results in a total of 203,616 diseased windows and 21,718 healthy windows. We highlight a target class imbalance in the dataset at two levels: 1) recording-level imbalance ratio of $\sim$7:1 (diseased:healthy) and 2) window-level imbalance ratio of $\sim$9:1 (diseased:healthy). The handling of imbalance at the window-level (relevant for ML training) is discussed later in the manuscript.

\noindent \textbf{Features:} The frequency content of the windowed EEG signals, obtained through the Power Spectral Density (PSD), was summarized into the dominant brain wave bands defined as follows: delta (1-4Hz), theta (4-7.5Hz), alpha (7.5-13Hz), lower beta (13-16Hz), higher beta (16-30Hz), and gamma (30-40Hz). We extracted the total band power from each band for each of the 8 montage channels, leading to a feature matrix of shape (8 channels x 6 features) for each window, as shown in Figure \ref{fig:feature_extraction}. Figure \ref{fig:boxplots} illustrates the differences between the two classes based on the extracted features using box-plots. Note that the features from channels in left and right hemispheres were averaged to generate combined features for each region. In addition, the features were z-scored for visual clarity.

\begin{figure}[h]
\floatconts
  {fig:boxplots}
  {\caption{Box-plots representing the group differences of spectral features.}}
  {\centering\includegraphics[width=0.9\linewidth]{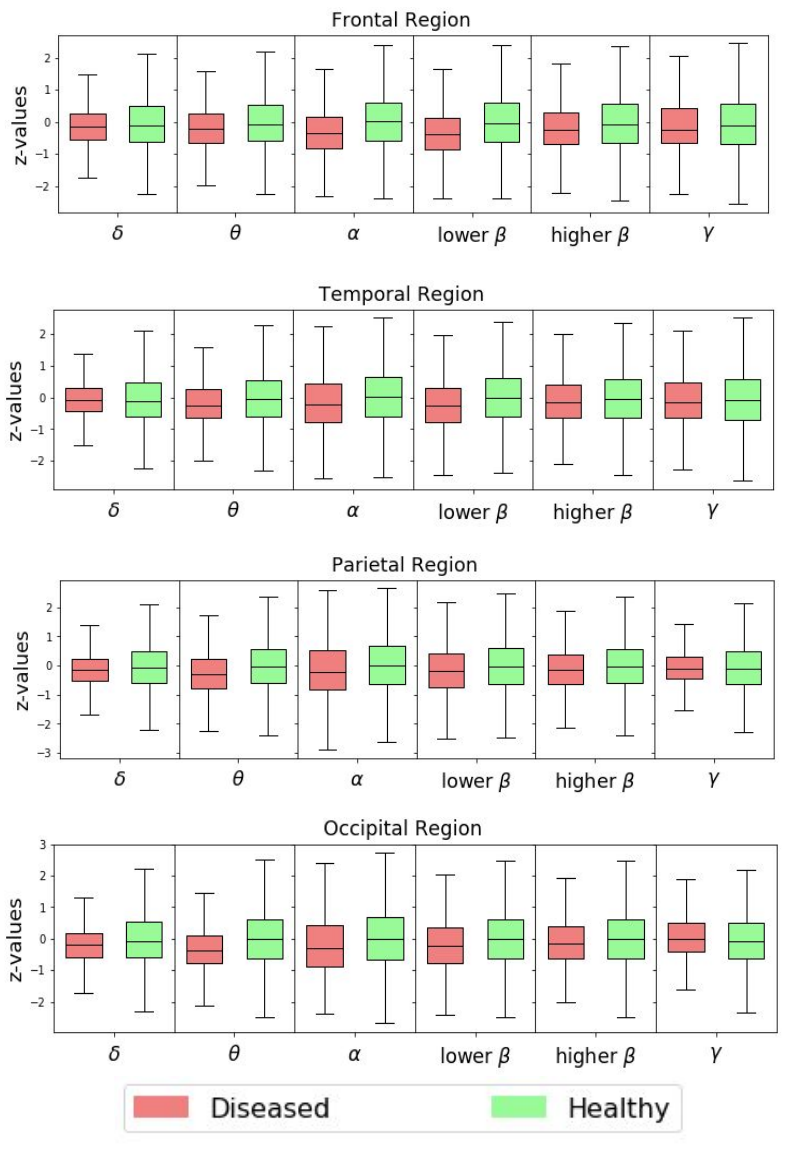}}
\end{figure}
\vspace{-10pt}


\section{Model Description}
\label{sec:meodel_desc}
In this section, we describe the motivation behind EEG-GCNN and mathematically formulate some of its distinct features. Overall, the aim of EEG-GCNN is to learn representations of EEG activity in a way that incorporates functional coupling (coordinated-firing activity) of distributed regions of the brain and the structural coupling (through physical white-matter tracts) between distant brain regions. 


Our proposed model (shown in Figure \ref{fig:eeg-gcnn}) consists of several parts. First, it describes the 10-second EEG window using a graph structure, which is given as input to the GCNN model. Second, graph convolutions are performed on the input data to generate node-level embeddings. Third, an averaging operation is performed across the nodes to generate graph-level embeddings. The graph-level embeddings are then provided as inputs to the fully-connected network, which predicts the output class. We describe each of these steps in the following sections.


\begin{figure*}[htbp]

\floatconts
  {fig:eeg-gcnn}
  {\caption{Graph representation of EEG data and the EEG-GCNN model architecture.}}
  {\centering\includegraphics[width=\linewidth]{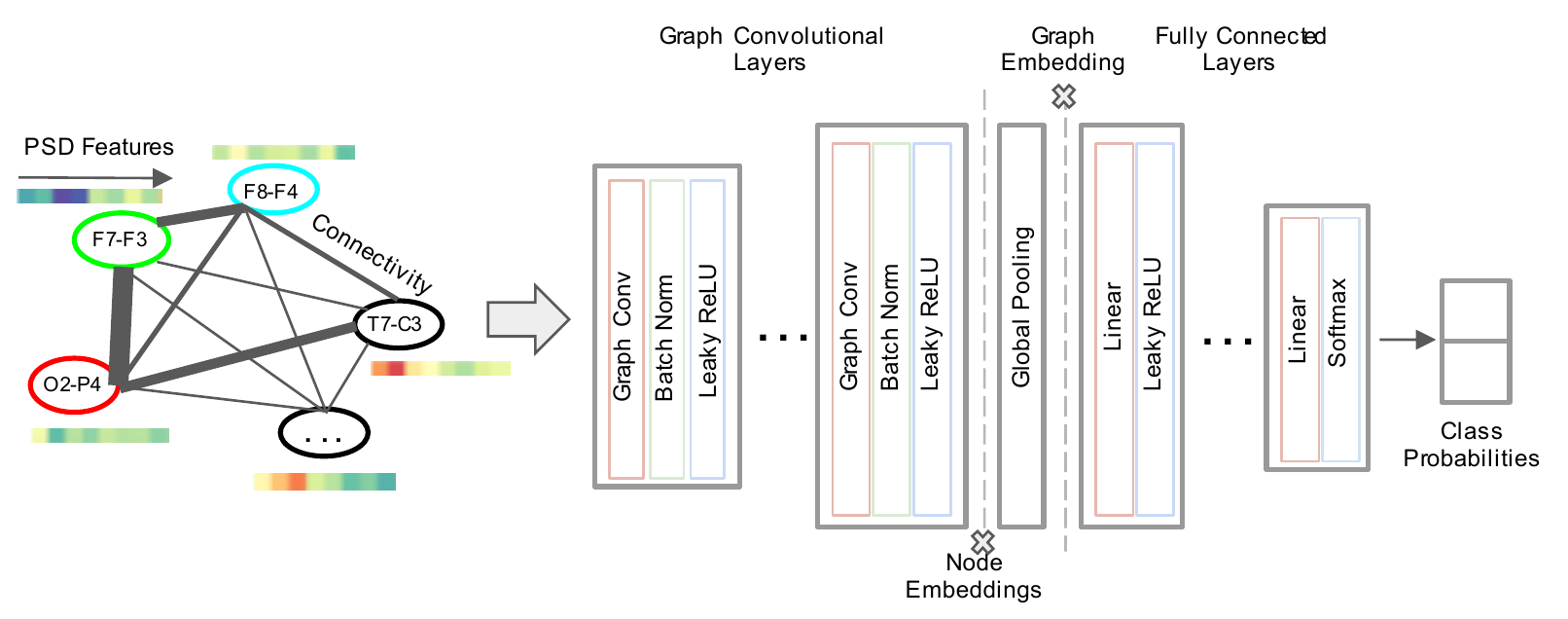}}
\end{figure*}



\noindent \textbf{Graph structure:} Suppose that EEG data of a subject are recorded through $M$ channels. Initially, the data is discretized by dividing the recording duration into $N$ windows. We represent the interactions between the channels at a window $n$ as a dynamic graph $ G_n = \left( V, E_n\right) $, where $V$ is the set of $|V| = M$ channels and $E_n \subset V \times V$ is the set of undirected links between channels. Data on $G_n$ can be represented by a feature matrix $X_n \in \mathbb{R}^{m\times d}$, where d denotes the input feature dimension per channel. The edge set E can be represented by a weighted adjacency matrix $A \in \mathbb{R}^{m\times m}$. 

\noindent \textbf{Connectivity measures:} The adjacency between channels $i$ and $j$ is denoted as $A_{ij}$ and is a combination of spatial ($A_{ij}^s$) and functional ($A_{ij}^f$) connectivity.
\begin{equation}
\label{eq:connectivity}
A_{ij} = \frac{1}{2}\left( A_{ij}^s + A_{ij}^f \right)
\end{equation} 

Specifically, the edge weight between each pair of nodes is made of two components: 1) the geodesic distance between the two electrodes when the standard 10-20 electrode configuration is mapped to a unit sphere (a proxy for spatial brain connectivity) and 2) the coherence values between the timeseries signal of the two electrodes (a proxy for functional brain connectivity). While the coherence values naturally lie in $[0, 1]$, the geodesic distances were standardized into the same range. Both these measures are then averaged to generate edge weights that lie within $[0, 1]$.

The geodesic distance $A_{ij}^s$ between two points on a sphere of radius $r$ with coordinates $(x_i, y_i, z_i)$ and $(x_j, y_j, z_j)$ in Cartesian coordinate space is defined as:
$$A_{ij}^s = \text{arccos}\left(\frac{x_ix_j + y_iy_j + z_iz_j}{r^2}\right)$$
The spectral coherence $A_{ij}^f$ between the two channel timeseries $i$ and $j$, with cross-spectral density $S_{ij}$ and power spectral densities $S_{ii}$ and $S_{jj}$, is defined as:
$$A_{ij}^f =  \frac{|\ E[S_{ij}]\ |}{\sqrt{E[S_{ii}]. E[S_{jj}]}}$$

\noindent \textbf{Graph convolutions:} We use the spectral graph convolution propagation rule defined by \citep{kipf2016semi}. Suppose there are $L$ number of graph convolutional layers and let $l=0,1,\dots,L-1$ denote the layer number. Each graph convolution produces a feature transformation of its inputs as described below, where $W^{(l)}$ denotes the weights of layer $l$, $H^{(l)}$ denotes the output of layer $l$ (with $H^{(0)} = X_n$), and $D$ denotes the diagonal degree matrix of graph $G_n$. Note that the degree matrix is trivial in our case because the graph is fully connected.
\begin{equation}
\label{eq:graph_conv}
H^{(l+1)} = \sigma\left( \hat{D}^{-\frac{1}{2}}\hat{A}\hat{D}^{-\frac{1}{2}}H^{(l)}W^{(l)}\right)
\end{equation} 
EEG-GCNN is aimed at graph-classification and therefore, node embeddings ($H^{(L)}$) are aggregated at the end of graph convolutions to form an embedding of the whole graph.

\noindent \textbf{Window-level predictions:} The graph embedding is then provided as input to a fully-connected network which produces output $Y_n \in [0,1]$, which represents the probability that the $n^{\text{th}}$ window was recorded from a patient with ND.

\noindent \textbf{Deriving subject-level predictions:} To estimate whether an entire EEG recording was recorded from a patient, we use a maximum likelihood estimation based on the window-level predictions. We model the window-level predictions of a subject $S_i$ as independent observations made from a Bernoulli trial with an unknown probability $\pi_i$, where $\pi_i$ is the probability that the subject $S_i$ is a patient. Then, an estimate of $\pi_i$ that maximizes the likelihood function $\prod_{n=1}^{N(i)}\pi_i^{Y(n)}(1-\pi_i )^{(1-Y(n))}$
after $N$ windows is given as 
$\hat{\pi_i} = \frac{\sum_{n=1}^{N}Y_n}{N}$.

\section{Evaluation Setup}

\begin{figure}[ht]
\floatconts
  {fig:cv_flow}
  {\caption{Evaluation procedure. We ensured that the data in training, validation, and testing sets were from disjoint sets of subjects.}}
    {\centering\includegraphics[width=\linewidth]{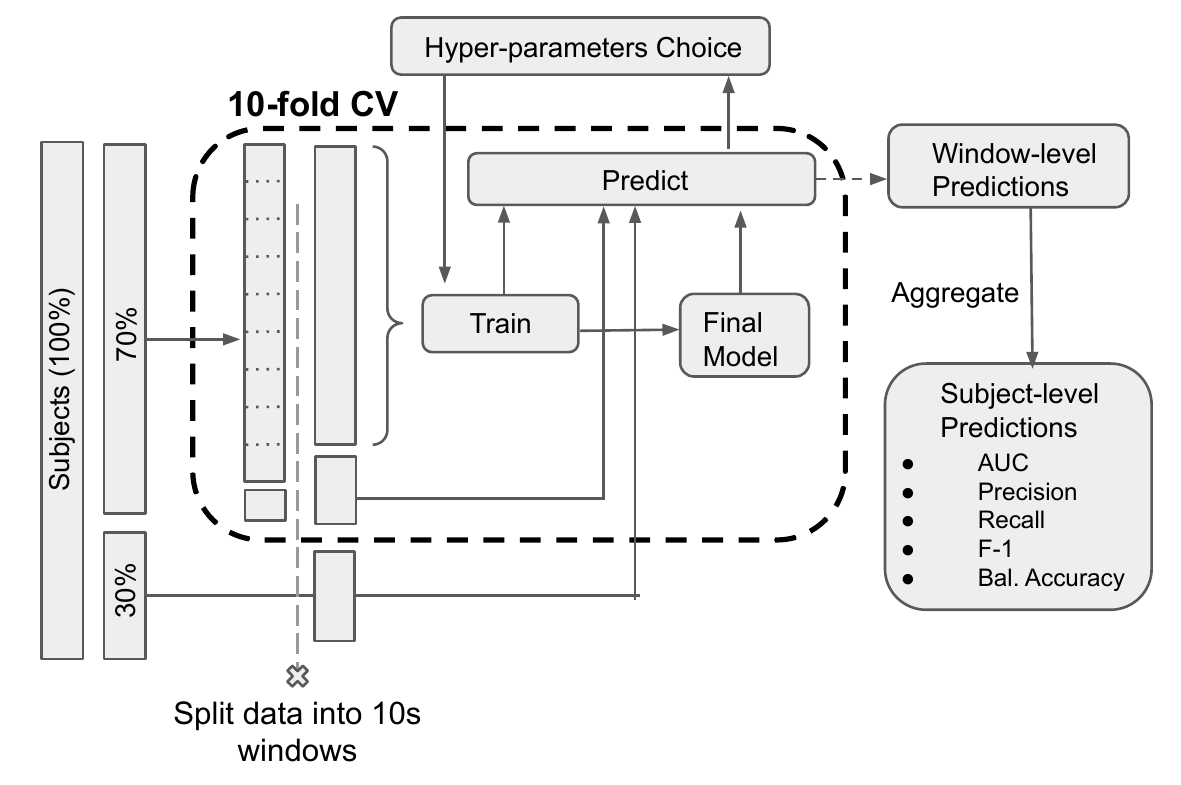}}
\end{figure}

\begin{table*}[ht]
    %
    \centering
    \begin{tabular}{l|cc}
        \hline
        Level & Train+Val (70\%) & Test (30\%)\\
        \hline
        Subjects & 1,115	(P: 964, H: 151) & 478 (P: 421, H: 57)\\
        Windows & 156,556 (P: 140,841, H: 15,715) & 68,778 (P: 62,775, H: 6,003) \\\hline\hline
    \end{tabular}
    \caption{Training, validation, and test sets. Abbreviations: P -- Patient, H -- Healthy.\label{tab:data}}%
\end{table*}


The evaluation procedure is depicted in Figure \ref{fig:cv_flow}. Table \ref{tab:data} summarizes divisions into training, validation, and testing sets. A 10-fold cross-validation (CV) routine was applied to the training set for each model to get robust estimates of model performance on validation and test sets. Note that each of the 10 CV folds maintained a disjoint set of participants in the resulting train and validation sets. This was done to ensure the validation performance reflects the performance on unseen subjects. Additionally, each model we compared was trained using the exact same 10 CV folds. We note that, a) the training and prediction are done at window-level, b) window-level predictions are aggregated to obtain subject-level predictions, and c) all CV folds share the same hyperparameters.



\noindent \textbf{Implementation of ML models:} We implemented several ML models for comparison as described in Section \ref{sec:experiments}. The shallow variant of EEG-GCNN comprises only of 2 Graph Convolution layers (output dimensions: 64, 128) and a Global Mean Pooling layer. Notably, the shallow EEG-GCNN has no hidden Linear layers. In contrast, the deep variant of EEG-GCNN, is composed of 5 Graph Convolution layers (output dimensions: 16, 16, 32, 64, 128), a Global Mean Pooling layer, and 2 hidden Linear layers (hidden dimensions: 30, 20). The baseline Fully-connected Neural Network comprises 2 hidden Linear Layers (hidden dimensions: 64, 32). The Random forest baseline comprises 100 decision tree learners, with 4 features considered for node splitting, and a maximum tree depth of 15. The bagging fraction was set to 20\% of the training set, and the complexity parameter for tree pruning was set to 0.015. For random forests, the predicted class probability of a sample is computed as the fraction of samples of that class present in the leaf node, averaged over all the trees in the forest. Model implementations were done using the Scikit-learn, PyTorch, and PyTorch Geometric libraries (\cite{pedregosa2011scikit}, \cite{paszke2019pytorch}, \cite{fey2019fast}). The models were trained on a Linux server with 64GB memory and two NVIDIA Titan Xp GPUs.

\noindent \textbf{Handling class imbalance:} The imbalance in the training data was handled by using a weighted cross-entropy loss function for the neural network models and evaluating a weighted Gini impurity score during random forests tree construction. The class weights were set to the inverse of the window count for that class, leading to a higher loss value for mistakes on the minority class.

\noindent \textbf{Evaluation metrics:} Model generalization performance was evaluated based on the receiver operating characteristic (ROC) curve made from subject-level class probabilities (calculated by averaging window-level probabilities). We report the average AUC scores obtained across the 10 CV folds. Additionally, we pick an optimal decision threshold using Youden's J statistic (\cite{youden1950index}) (i.e., the threshold that maximizes the sum of sensitivity and specificity) and use it to report the precision, recall, F-1, and balanced accuracy scores. While precision, recall, and F-1 scores are calculated treating the ``patient" EEGs as the positive class, balanced accuracy is calculated as the average of the recall of each class.

\subsection{Hyperparameter Tuning}

In the following, we highlight the heuristics and strategies used in the tuning process. The exact values of the hyperparameters chosen for the final models can be found in the model definition files of released software.

\noindent \textbf{Neural networks:} The use of batch normalization made training vastly more robust to specific choices of hyperparameters. Nonetheless, layers with more trainable parameters required stronger dropout regularization. Layer depth was increased until performance gains were marginal. The initial learning rate of Adam optimizer was set to 0.1 and then decayed by a factor of 10 at regular intervals. The training process was visualized extensively to determine saturation and identify less favorable settings.

\noindent \textbf{Random forests:} Random forests was tuned through 3 rounds of iterative grid search (1512 configurations) on a fixed validation set. Starting from a wide uniform grid, next rounds focused on favorable grid regions. Performance was seen to be sensitive to 5 variables: total estimators in the forest, the maximum depth of tree, the cost-complexity pruning parameter, the bagging fraction, and the minimum number of samples at leaf nodes. The best performing configuration was then used in 10-fold CV.

\section{Experiments \& Results}
\label{sec:experiments}

\begin{table*}[htbp]
    %
    \resizebox{\linewidth}{!}{
    \begin{tabular}{l|ccccc}
        \hline
        Model & AUC & Precision & Recall & F-1 & Bal. Accuracy \\
        \hline
        FCNN & 0.71 (0.08) & 0.94 (0.02) & 0.66 (0.11) & 0.77 (0.08) & 0.66 (0.07)\\
        Random Forests & 0.80 (0.01) & 0.95 (0.01) & 0.79 (0.08) & 0.86 (0.05) & 0.74 (0.02)\\
        \textbf{Deep EEG-GCNN} & \textbf{0.90 (0.04)} & 0.99 (0.00) & 0.74 (0.08) & 0.84 (0.06) & \textbf{0.85 (0.04)}\\
        \textbf{Shallow EEG-GCNN} & \textbf{0.90 (0.02)} & 0.99 (0.01) & 0.72 (0.07) & 0.83 (0.04) & 0.83 (0.02)\\
        \hline
        Trivial Classifier 1 & 0.50 (0.02) & 0.88 (0.01) & 0.87 (0.02) & 0.87 (0.01) & 0.50 (0.02)\\
        Trivial Classifier 2 & 0.50 (N/A) & 0.88 (N/A) & 1.00 (N/A) & 0.94 (N/A) & 0.50 (N/A)\\\hline\hline
    \end{tabular}
    }
    \caption{Results on the held-out set of 478 subjects. All metrics were calculated at the subject-level treating patient EEGs as the positive class. N/A indicates no variability.\label{tab:results}}%
\end{table*}

\noindent \textbf{Model comparisons:} 
We employ the evaluation procedure discussed above to compare the performance of a shallow and deep EEG-GCNN architecture against two classical ML baselines - fully-connected neural networks (without graph convolutions) and random forests. The results of this comparison on held-out subjects are presented in Table \ref{tab:results} and Figure \ref{fig:results} displays the corresponding ROC curves. Our results indicate that GCNN-based approaches outperform both the ML baselines ($\sim10\%$ improvement in AUC and balanced accuracy). Additionally, a Kolmogorov-Smirnov test between GCNN models and ML baselines rejected the null hypothesis that the subject-level probabilities (averaged across CV folds) are drawn from the same distribution ($p<0.05$). This finding suggests that the differences in AUC are statistically significant. However, based on the optimal threshold chosen, our results indicate that random forests provides improved recall and F1 score despite providing lower AUC and precision.

\noindent To highlight chance-level performance and provide additional context for interpreting model comparisons in light of imbalanced data, we provide the performance of two trivial classifiers. We consider two blind classifiers i.e., the classifiers that were not trained with data: 1) a classifier that predicts the positive class with the label imbalance probability of $\sim$0.86, whose results are reported after 1000 simulations (referred to as trivial classifier 1) and 2) a classifier that always predicts the majority  class (i.e., patient) regardless of the input (referred to as trivial classifier 2). While trivial baselines achieve a balanced accuracy and AUC of 0.50, the ML models, particularly EEG-GCNN, shows a marked improvement over all baselines.

\begin{figure}[ht]
\floatconts
  {fig:results}
  {\caption{The mean ROC curves, where the shaded region indicates 95\% confidence intervals.}}
  {\centering\includegraphics[width=\linewidth]{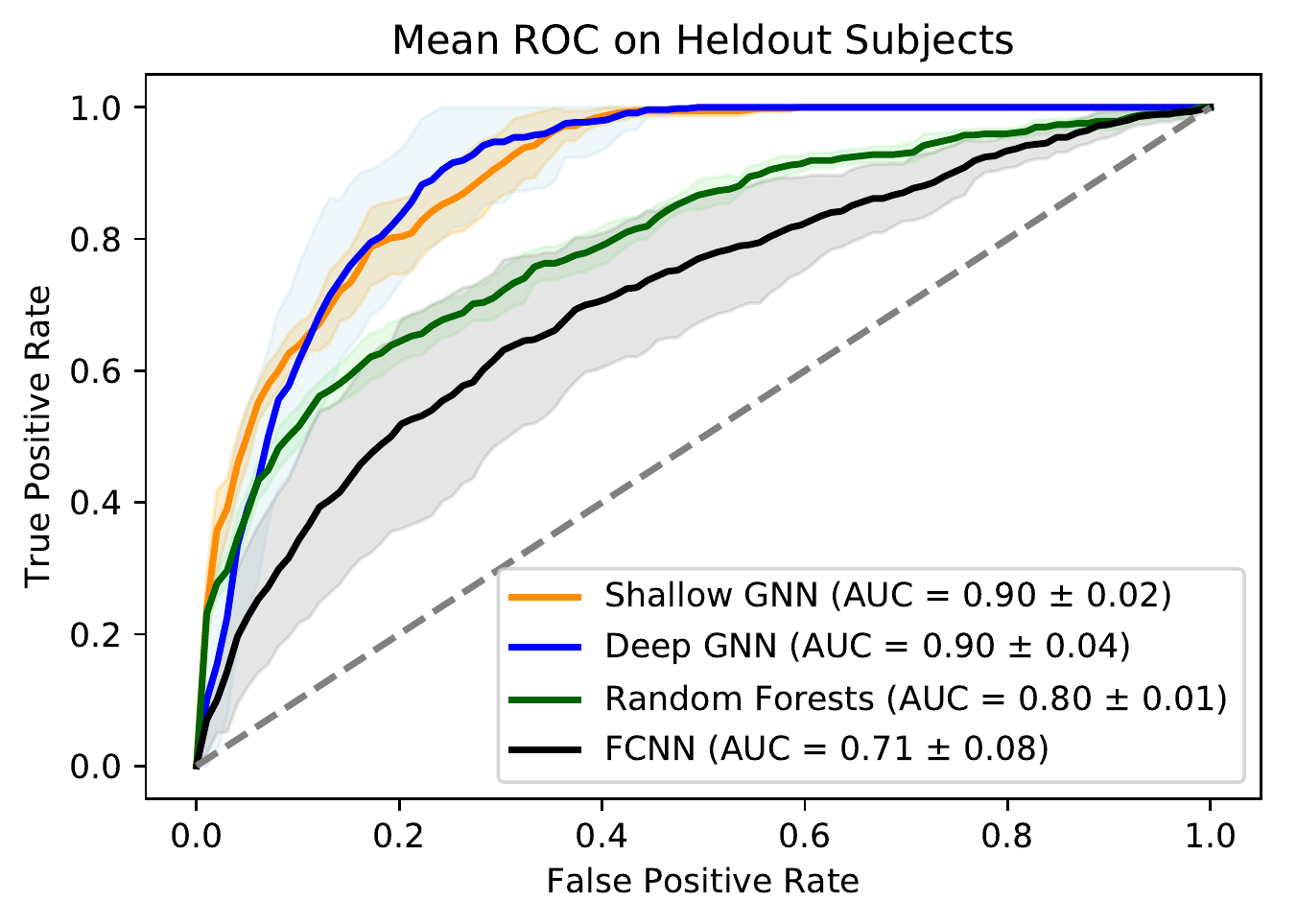}}
\end{figure}

\noindent \textbf{Shallow or Deep Model:} We evaluated the effect of GCNN network depth on performance with shallow and deep EEG-GCNN variants. Our results suggest that larger depth provides only a marginal improvement in performance over shallower variants. This observation is in line with multiple previous GCNN studies that report marginal reductions in performance with increasing network depth beyond just 2-4 layers \citep{zhao2019pairnorm}. 


\noindent \textbf{Large vs small training size:}
We observed an additional increase in performance of shallow EEG-GCNN when trained on only a tenth of the training data. This result is counter-intuitive in the deep learning paradigm and goes against conventional wisdom of increasing performance by using more data. However, a recent study \citep{nakkiran2019deep} shows that shallower models perform worse on larger data for a certain range of model complexity. While FCNN showed characteristics of overfitting, both shallow and deep EEG-GCNN models did not. Deep GNN remains unaffected while FCNN shows a drop in performance, as shown in Table \ref{tab:datasize}.

\begin{table*}[ht]
    %
    \centering
    \begin{tabular}{l|ccc}
        \hline
        Model & Train & Val & Test\\
        \hline
        FCNN & 0.80 (0.08) & 0.64 (0.03) & 0.67 (0.07)\\
        Shallow EEG-GCNN & 0.96 (0.03) & 0.94 (0.02) & \textbf{0.92 (0.03)} \\
        Deep EEG-GCNN & 0.93 (0.04) & 0.92 (0.03) & 0.90 (0.05) \\\hline\hline
    \end{tabular}
    \caption{Training, validation, and test AUCs when trained using only a tenth of the data.\label{tab:datasize}}%
\end{table*}





\begin{figure}[ht]
\floatconts
  {fig:embeddings}
    {\caption{t-SNE maps of heldout test subject embeddings. Electronic zoom recommended for viewing. Green denotes a healthy subject, while red denotes a diseased patient. (a) 2D map of the embeddings from all 10 cross-validation models plotted together. (b) 3D map of one cross-validation fold.}}
  {\centering \includegraphics[width=\linewidth]{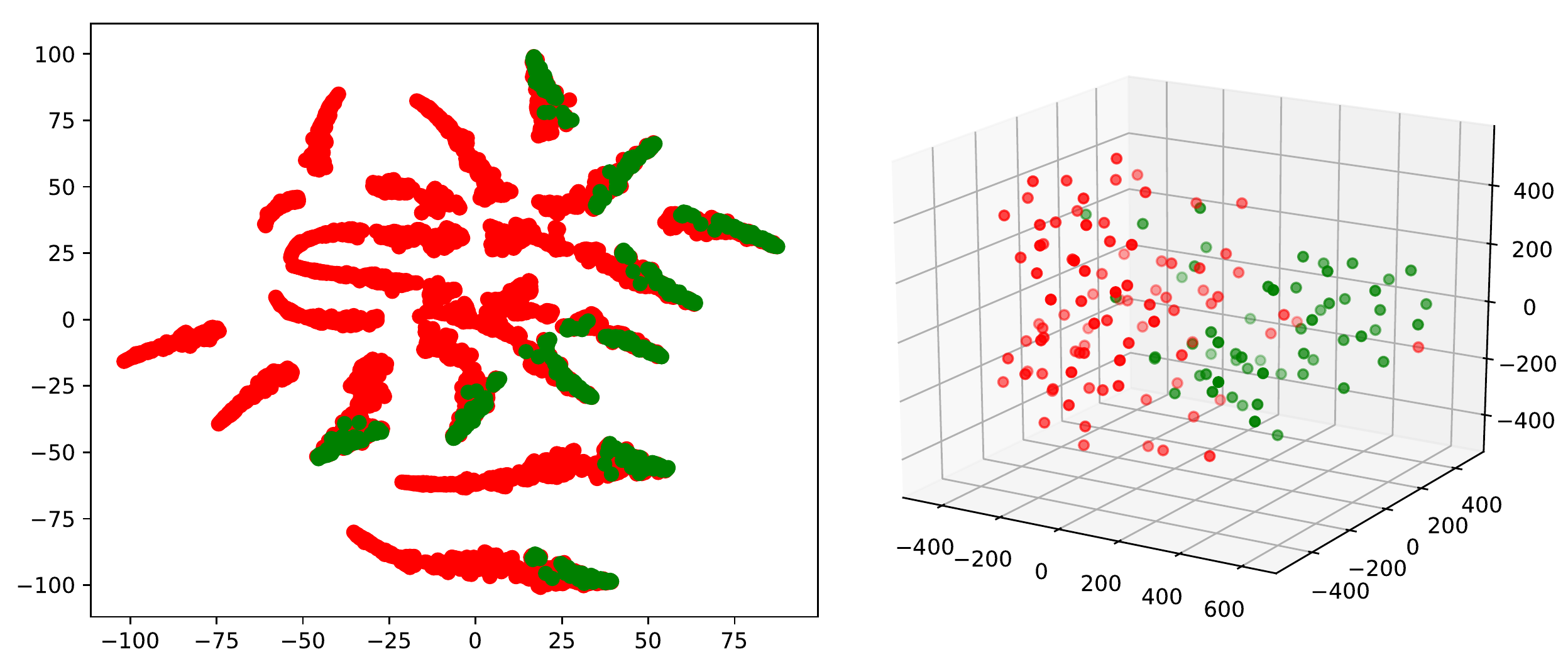}\hfill
    \centering \includegraphics[width=\linewidth]{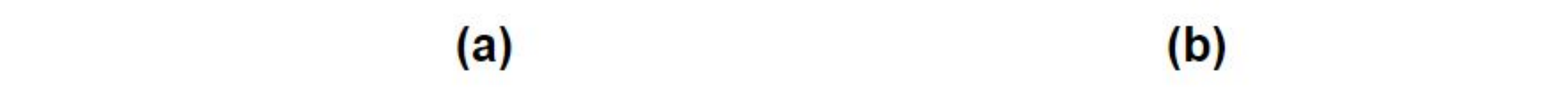}\hfill}
\end{figure}

\noindent \textbf{Improved linear separability:} Spectral features show faint distributional clues, if any, (Figure \ref{fig:boxplots}) while shallow EEG-GCNN (AUC: 0.90) dramatically outperforms a vanilla FCNN (AUC: 0.71). This is interesting because shallow EEG-GCNN does not include any hidden layers in the fully-connected part, suggesting that graph convolutions learn better feature representations. We perform a qualitative assessment of the 128-dimensional EEG-GCNN embeddings using the t-SNE algorithm (\cite{maaten2008visualizing}), results of which are shown in Figure \ref{fig:embeddings}. Note that each point in the scatter plot is a subject (red denotes patient, green denotes healthy). Figure \ref{fig:embeddings}a shows the t-SNE maps of the embeddings from all 10 CV models plotted together, i.e., each subject is plotted 10 times, and Figure \ref{fig:embeddings}b shows the t-SNE maps of the embeddings of a single CV fold. We find that the EEG-GCNN embeddings, in general, show better separability compared to spectral features.

\vspace{-10pt}
\section{Discussion \& Future Work}
Conventional analysis of EEG relies on expert annotations of various phenomena (e.g., awake and sleep, artifacts, bad channels). Such annotations are time consuming, costly, susceptible to human error, and clearly not scalable. We presented a fully-automated approach based on graph neural networks that does not require expert annotations of specific events. As such, our approach provides multiple benefits: 1) it enables large scale studies, 2) it can eliminate individual biases, and 3) it can augment the visual review of EEGs by providing focused inputs and help reduce physician burnout \citep{verghese2018computer}. Our future efforts will focus on developing end-to-end models with raw data/spectrograms as inputs and implementing saliency methods, as they can help identify novel EEG features and advance neurological disease research. 

A limitation of our study is that the EEGs of two populations, healthy and patients, were acquired using different systems under different conditions. Therefore, the difference between the acquisition systems/environments is a potential confounder in our analyses. To address this limitation as best we could, we undertook the same preprocessing steps for both the EEG datasets. Regardless, future studies including EEGs of both controls and patients recorded using the same acquisition system are necessary to eliminate this confounder and to elucidate the clinical value of our approach.

Furthermore, scalp-EEG provides a rich representation of the underlying brain state with substantial spatial and temporal granularity. However, the presence of artifacts makes the decoding of underlying brain state a nontrivial task. We postulate that the  representation of EEG data learned by our model can be used to describe the underlying brain state and has potential utilities in other areas of brain research such as sleep staging, brain computer interfaces, and neural state decoding. In the future, we will also investigate the possibility of classifying various brain states (e.g., sleep stages) using the core methodology developed in this study.


\noindent \textbf{Code \& data availability:} The datasets used in this study are already publicly available. The final trained models, data set metadata, and code to reproduce Table \ref{tab:results} results are available at \url{https://github.com/neerajwagh/eeg-gcnn}.


\section{Conclusion}
We introduce EEG-GCNN, a novel GCNN architecture for multi-channel EEG data inspired by the clinical significance of brain connectivity patterns in neurological disease pathology. We apply EEG-GCNN on a large dataset of ``normal" EEG recordings from 1593 subjects and present strong evidence for the ability to distinguish between ``normal" EEGs of neurologically diseased individuals and the EEGs of healthy individuals (AUC: 0.90). Clinical use of the superior predictive capability of EEG-GCNN can shorten the traditional diagnosis process and help expert neurologists make more accurate diagnoses.



\section{Acknowledgements}

We would like to thank our Mayo Clinic collaborators Gregory A. Worrell, Benjamin H. Brinkmann, and Brent M. Berry, for providing neurological domain expertise and validating the clinical significance of our proposed hypothesis. In addition, we would like to thank the Mayo Clinic Neurology Artificial Intelligence Initiative and the Mayo Clinic Illinois Alliance for Technology-based Healthcare Research, for providing financial and logistical support for this research.

\bibliography{main_paper}

\end{document}